\documentclass[unnumsec,webpdf,contemporary,large]{oup-authoring-template}%
\usepackage{ulem}
\graphicspath{{Fig/}}

\theoremstyle{thmstyleone}%
\theoremstyle{thmstyletwo}%
\theoremstyle{thmstylethree}%

\newsavebox\CBox
\def\textBF#1{\sbox\CBox{#1}\resizebox{\wd\CBox}{\ht\CBox}{\textbf{#1}}}

\begin{document}

%%%%%%%%
\journaltitle{Briefings in Bioinformatics}
\DOI{DOI HERE}
\copyrightyear{2023}
\pubyear{2023}
\access{Advance Access Publication Date: Day Month Year}
\appnotes{Paper}
%%%%%%%%
\firstpage{1}

%\subtitle{Subject Section}

% \title[Exploring Peptide Encoding with Deep Learning]{Exploring Peptide Encoding with Deep Learning through A Sizeable Peptide Self-assembly Dataset}

\title[Efficient Prediction of Peptide Self-assembly]{Efficient Prediction of Peptide Self-assembly through Sequential and Graphical Encoding}

\author[1,2$\ast$]{Zihan Liu\ORCID{0000-0001-6224-3823}}
\author[3,4$\ast$]{Jiaqi Wang}
\author[1,4]{Yun Luo}
\author[3,4]{Shuang Zhao}
\author[3,4$\dagger$]{Wenbin Li}
\author[2$\dagger$]{Stan Z. Li}

\authormark{Zihan Liu et al.}

\address[1]{\orgdiv{College of Computer Science and Technology}, \orgname{Zhejiang University}, \orgaddress{Hangzhou, \country{China}, \postcode{310058}}}
\address[2]{\orgdiv{AI Lab, Research Center for Industries of the Future}, \orgname{Westlake University}, \orgaddress{Hangzhou, \country{China}, \postcode{310030}}}
\address[3]{\orgdiv{Research Center for the Industries of the Future}, \orgname{Westlake University}, \orgaddress{Hangzhou, \country{China}, \postcode{310030}}}
\address[4]{\orgdiv{School of Engineering}, \orgname{Westlake University}, \orgaddress{Hangzhou, \country{China}, \postcode{310030}}}

\corresp[$\ast$]{Co-first authors: \href{liuzihan@westlake.edu.cn}{liuzihan@westlake.edu.cn}, \href{wangjiaqi@westlake.edu.cn}{wangjiaqi@westlake.edu.cn}\;}

\corresp[$\dagger$]{Corresponding authors: \href{stan.zq.li@westlake.edu.cn}{stan.zq.li@westlake.edu.cn}
\href{liwenbin@westlake.edu.cn}{liwenbin@westlake.edu.cn}}

\received{Date}{0}{Year}
\revised{Date}{0}{Year}
\accepted{Date}{0}{Year}

\abstract
{In recent years, there has been an explosion of research on the application of deep learning to the prediction of various peptide properties, due to the significant development and market potential of peptides. 
Molecular dynamics has enabled the efficient collection of large peptide datasets, providing reliable training data for deep learning.
However, the lack of systematic analysis of the peptide encoding, which is essential for AI-assisted peptide-related tasks, makes it an urgent problem to be solved for the improvement of prediction accuracy.
To address this issue, we first collect a high-quality, colossal simulation dataset of peptide self-assembly containing over 62,000 samples generated by coarse-grained molecular dynamics (CGMD).
Then, we systematically investigate the effect of peptide encoding of amino acids into sequences and molecular graphs using state-of-the-art sequential (i.e., RNN, LSTM, and Transformer) and structural deep learning models (i.e., GCN, GAT, and GraphSAGE), on the accuracy of peptide self-assembly prediction, an essential physiochemical process prior to any peptide-related applications.
Extensive benchmarking studies have proven Transformer to be the most powerful sequence-encoding-based deep learning model, pushing the limit of peptide self-assembly prediction to decapeptides.
In summary, this work provides a comprehensive benchmark analysis of peptide encoding with advanced deep learning models, serving as a guide for a wide range of peptide-related predictions such as isoelectric points, hydration free energy, etc.
}


\maketitle

\section{Introduction}

Peptides are miniature proteins typically consisting of less than 50 amino acids \cite{langel2009introduction}.
A variety of advantageous properties such as mature and facile synthesis, biocompatibility and biodegradability, and rich chemical diversity make self-assembly peptides have a wide range of applications \cite{cinar2012amyloid,fan2018near,tao2017self,zhao2010molecular}. 
The self-assembly of peptides have been implicated in various biological processes \cite{levin2020biomimetic,whitesides2002self,krause2014steering} and protein misfolding diseases \cite{aguzzi2009prions,knowles2014amyloid,chiti2017protein}.
The elucidation of the self-assembly mechanisms with respect to the type and position of 20 natural amino acids should be crucial for the efficient design of peptide self-assemblies and contribute to drug discovery and, thus, to public health. 

Since only a small number of self-assembled peptides have been found so far within the huge sequence space ($>$$20^{50}$), in order to accelerate the discovery of self-assembling peptides, computational screening employing coarse-grained molecular dynamics (CGMD) \cite{rudd1998coarse} has been adopted.
Compared to the design of wet-lab experiments, computational screening of CGMD with a reliable force field can significantly accelerate the discovery process due to parallel computing.
Dating back to 2014, Frederix et al. \cite{frederix2015exploring} explored the complete sequence space of tripeptides with 8,000 sequence quantities, and a series of design rules were first proposed and validated by experiments.
For longer peptides, it becomes an increasing challenge to explore the complete sequence space using CGMD due to the high computational cost incurred by exponentially growing sequence quantities.
To address this challenge, Batra et al. \cite{batra2022machine} used machine learning (ML) coupling CGMD to search for the self-assembling sequences within the entire sequence space of pentapeptides.
They generated training samples by simulation and predicted the aggregation tendency of other peptides in the space using random forest and Monte Carlo tree search.
Due to the huge computational cost and time involved, it is hard to increase the size of the simulation data with the exponential expansion of the peptide sequence space with sequence length.
Therefore, the development of AI models to assist simulated data has become the key to research.

This work aims to comprehensively evaluate and analyse advanced deep learning modelling methods through peptide encoding directions of amino acid sequences and molecular graphs.
To evaluate deep learning methods, it is necessary to have a large dataset with low noise.
Therefore, CGMD is used to generate data for peptide sequences using aggregation propensity (AP) as a label.
It should be noted that the self-assembly of peptides involves two steps, firstly, peptides aggregate into oligomers and secondly self-assemble into supramolecular structures \cite{zapadka2017factors}. Thus, the aggregation of peptides is a prerequisite of self-assembly. 
In this work, more than 62,000 samples (i.e., one sample for a peptide sequence and its corresponding AP) have been collected from the full space of pentapeptides to decapeptides, costing over \$50,000.
To the best of our knowledge, this is the largest dataset of peptide self-assembly simulations.
Based on the large, low-noise dataset, deep learning modelling is then performed from the perspective of peptide encoding.
Due to the short length of single peptide, its conformation is relatively simple before aggregation \cite{marullo2013peptide,seebach2006helices}. It is typically a random coil, not as complex as the three-dimensional structures of its aggregates (i.e., extend $\beta$-sheet).
Therefore, the embedding of the primary structure is crucial for peptide-related tasks.
Previous work \cite{batra2022machine} encoded the peptides as one-dimensional vectors, which is unable to reveal the structural information of the peptide.
To better extract the structural information of peptides by AI, we propose sequence encoding based on the amino acid sequence and graph encoding based on the molecular graph.
Sequence encoding encodes a peptide as an amino acid sequence with alignments, and the sequential representation of a peptide is then extracted by a sequence model such as Long Short-Term Memory (LSTM) \cite{hochreiter1997lstm}, and Transformer \cite{vaswani2017attention}.
Graph encoding encodes a peptide as a molecular graph consisting of beads and chemical bonds, and Graph Neural Networks (GNNs) \cite{kipf2016semi,velivckovic2017graph,hamilton2017inductive} are employed to extract the graphical representation of the peptide.
To compare sequence-based and graph-based encoding with their modelling approaches, we evaluate rich advanced deep learning models on the self-assembly dataset.
We compare the performance of advanced models for 1D vector, sequence and graph encoding on the regression and classification tasks, respectively.
The encoding methods are further discussed and analysed in terms of model performance and applicability.
The dataset and the implementation of peptide encoding with advanced deep learning models involved in this work are open source to facilitate research on self-assembly and peptide-related tasks.

\section{Dataset Collection and Reliability Analysis}

\subsection{Selection of Peptide Samples}

Being aware of an enormous chemical space of $20^{10}$ decapeptides, Monte Carlo sampling is computationally unaffordable.
To diminish sampling bias, within the sequence space of pentapeptides to decapeptides, we sample over 62,000 peptide sequences using the Latin Hypercube \cite{mckay2000comparison} sampling approach. 
Latin hypercube sampling is a statistical technique that can be used to efficiently sample parameter spaces. 
When applied to peptide sampling, it can have several advantages over other sampling methods, such as increased diversity, reduced bias, efficient use of resources, and improved statistical analysis. In summary, Latin Hypercube sampling is an efficient and unbiased method for peptide sampling that can improve the quality of results while conserving resources.

\subsection{Aggregation Propensity (AP) Labelling}

\begin{figure}
    \centering
    \includegraphics[width=\hsize]{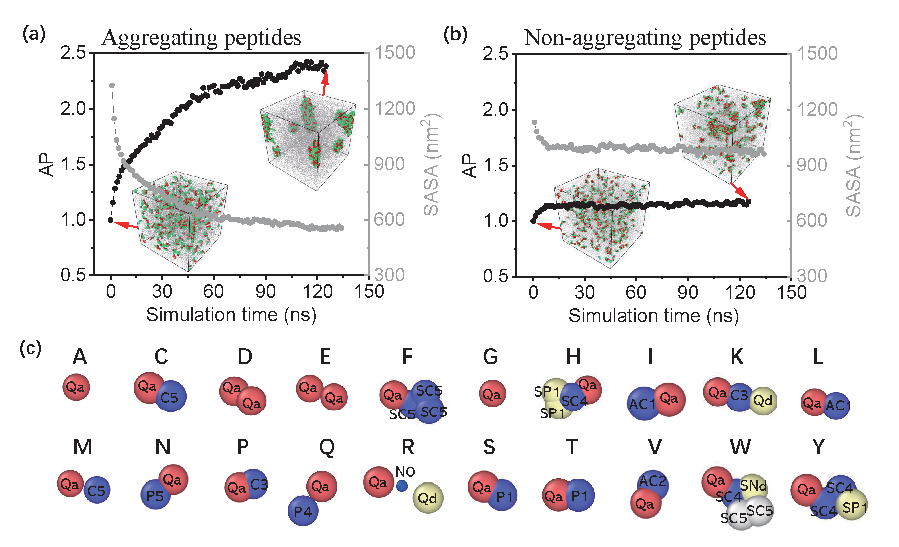}
    \caption{(a,b): Evolution of aggregation propensity (AP) and accessible surface area (SASA) within simulation duration of 125 ns. For aggregating peptides (a), the AP increases gradually due to the minimization of SASA; while for non-aggregating peptides (b), both AP and SASA remain a constant, and AP approximately equals 1, and an AP larger than 1.5 indicates an aggregating peptide. In the Fig. S1 of Supplementary file, we provide separate visualisations of the simulation process for an aggregated peptide and a non-aggregated peptide. (c) Bead representation of 20 natural amino acids.}
    \label{fig2}
\end{figure}

The sampled sequences are simulated using the CGMD simulations to generate a label termed Aggregation Propensity (AP), performed with the open-source package GROMACS \cite{abraham2015gromacs} and Martini force field \cite{marrink2007martini,monticelli2008martini}. 
The AP is defined as the ratio between the solvent accessible surface area (SASA) at the beginning (SASA$_{initial}$) and at the end (SASA$_{final}$) of a CGMD simulation \cite{frederix2015exploring}, as shown in the following equation:
\begin{equation} \label{AP}
    AP = \frac{SASA_{initial}}{SASA_{final}}
\end{equation}
At the beginning of the simulation, AP is equal to 1. 
As the simulation progresses, for aggregating peptide systems, the accessible surface area will decrease, and thus AP will increase above 1 (Figure \ref{fig2} (a)); for non-aggregating peptides, the accessible surface area will be approximately the same (due to calculation error), and AP will oscillate around 1 (Figure \ref{fig2} (b)).
This is not a serious criterion for judging whether a peptide is aggregating, as AP values can be affected by the duration of the simulation. 
However, within the same simulation, larger AP values indicate a greater tendency to aggregate, and normally, an AP larger than 1.5 indicates an aggregating peptide.

Prior to CGMD simulations, the all-atom peptide structures are generated based on CHARMM36 \cite{huang2013charmm36}, and then coarse-grained (CG) using the Python script martinize.py \cite{marrink2007martini,monticelli2008martini}. 
The CG model usually approximates 4 atoms (rarely 2, 3, or 5 atoms) into one bead in order to speed up simulations. Currently, eighteen particle types are defined, divided into four main categories: P, polar; N, intermediate polar; C, apolar; Q, charged. Each category has a number of sublevels (0, a, d, or da): subtype 0 has no hydrogen bonding capacity; subtype a has some hydrogen acceptor capacity; subtype d has some hydrogen donor capacity; subtype da has both donor and acceptor capacity; or (1, 2, 3, 4, 5) where subtype 5 is more polar than 1. 
S indicates that the bead is a ring-type bead. Standard beads are mapped using a 4:1 mapping scheme, and ring beads which are used for ring compounds are mapped 2-3:1 \cite{marrink2007martini,monticelli2008martini}. The bead representations of 20 natural amino acids are shown in Figure \ref{fig2}.
It should be noted that the backbone bead of each amino acid is different due to the different charge states at the N or C terminus, whereas the beads shown here are the single amino acid dissolved in water solvent as a zwitterion.

After the coarse-graining of the all-atom peptide structure, we placed multiple peptide molecules of the same type in simulations and observed their tendency to aggregate. A total of 150 coarse-grained pentapeptides (or 81 decapeptides) were randomly solvated in a 15 nm$\times$15 nm$\times$15 nm box with 28,400 water beads (water density is about 1 g/cm$^3$), resulting in a solvent concentration of 0.074 mol/L for pentapeptides (0.040 mol/L for decapeptides), close to that in the experiment. We repeated this solvation process for all the 62159 types of peptides to obtain the AP values.
The charge of the solution is kept neutral by adding an appropriate amount of Na$^+$ or Cl$^-$. The whole system is then energy-minimized using the steepest descent algorithm until the maximum force on each atom is less than 20 kJ$\cdot$mol$^{-1}$$\cdot$nm$^{-1}$.
The system is then subjected to an equilibration run for 5$\times$10$^6$ steps, with a time step of 25 fs, resulting in a total simulation time of 125 ns.
The temperature and pressure during the equilibration are controlled by the Berendsen algorithm at 300 K and 1 bar, respectively. 
The AP value is calculated in the last step using Equation \ref{AP}. 

\subsection{Dataset Quality}

For a computational framework coupling AI and CGMD
simulation, data quality should be described from two perspectives.

The first is the reliability of the CGMD simulations.
The most critical setting in CGMD simulation is the force field \cite{marrink2007martini,monticelli2008martini}, which affects the simulation accuracy and computational efficiency. 
One of the most prevalent force fields is the Martini force field, which has been widely adopted in peptide/protein self-assembly studies, and the proposed new physics have been confirmed by experiments \cite{lee2012modeling,xiong2019conformation,mueller2016machine}. 
It is therefore expected that the level of self-assembling can be well-represented in CGMD simulations and is adopted as well in this work.
A comparison of the AP values from our predictions derived from the training data generated by the CGMD and CGMD-derived AP values from available literature \cite{batra2022machine} has been shown in Tab. S1, Tab. S2 and Fig. S2 of Supplementary file. A good agreement between the prediction and the literature has been achieved, consolidating the reliability of the CGMD simulations and prediction models.

The second is the advantage of simulation dataset over real-world dataset for training AI models, which has been broadly discussed and validated \cite{halevy2009unreasonable,nikolenko2021synthetic,tremblay2018training}. (1) The simulation data has much more diversity and is much larger in size. The Latin hypercube approach that we used to sample the peptide sequences ensures that the frequency of individual amino acid at each position is consistent, i.e., the peptide samples are not biased to specific amino acids.  Also, the number of peptide sequences in our work (i.e., 62,159) is much larger than number of the tractable real-world samples (e.g., 100), considering the time and cost consumed in experiments. (2) The simulation data tend to have lower noise. This is because simulations are based on physical models, and their numerical errors are smaller and can be further reduced by increasing the simulation time or improving the simulation algorithms. Real-world samples, on the other hand, are affected by many factors, such as experimental conditions, instrument noise, and observation noise, leading to more variations and noise than the simulations.

\section{Methodology of Peptide Encoding and Modeling}

\begin{figure*}
    \centering
    \includegraphics[width=\hsize]{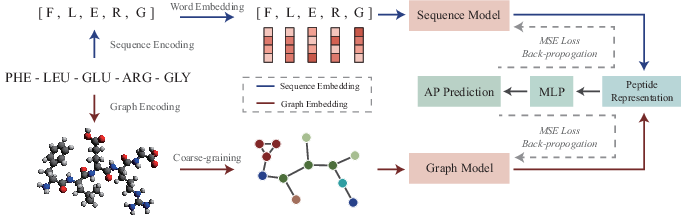}
    \caption{Framework diagram of peptides from encoding to modelling. The blue and red arrows indicate the sequence and graph encoding with modelling processes. A peptide undergoes encoding, extraction of representations, and classification by a decoder to obtain AP predictions. Predictions are supervised by the loss function, and the network is updated by back-propagation.}
    \label{fig1}
\end{figure*}

\subsection{Encoding Peptides in Amino Acid Sequences} \label{4.1}

Since peptides are sequence-like molecules formed by the dehydration-condensation reaction of amino acids, most studies have encoded peptides in the FASTA form, which are amino acid sequences.
The encoding and modelling of peptides into sequences are illustrated in Figure \ref{fig1}, indicated by the blue lines.
Each peptide is expressed as a sequence starting at the -NH$_2$ group.
As the length of the peptides in the dataset ranged from 5 to 10, peptides with lengths less than are made up by padding.
Each word in the sequence is mapped into a high-dimensional vector by learnable input embedding, same as in reference \cite{vaswani2017attention}, which can be understood as a one-hot encoding of the amino acid library and mapping into a high-dimensional space using a linear layer.
The encoded peptide is represented as a sequence of vectors $\{X_1,X_2,... ,X_N\}$, where $N$ is the length of the input peptide.
Each amino acid of the peptide sequence $X_t$ is represented as a vector $(x_1,x_2,...,x_d)$, where $d$ is the dimension of the input embedding.

We then employ sequential deep learning models, including Recurrent Neural Network (RNN) \cite{elman1990rnn}, LSTM \cite{hochreiter1997lstm}, Bi-LSTM \cite{schuster1997bilstm}, and Transformer \cite{vaswani2017attention}, to extract amino acid sequence representations.
The first three models belong to the RNN family.
Vanilla RNN suffers from exploding or vanishing gradients, resulting in its inability to learn long-term dependencies in sequences.
LSTM is proposed and has become a popular solution, while it only considers the previous context of the sequence.
Bi-LSTM, a bi-directional method based on LSTM, is proposed that considers both the previous and future information to estimate the representation.
Transformer \cite{vaswani2017attention} is one of the most advanced sequence model frameworks that has been widely used in recent years.
Unlike the recurrent models, it uses parallel input for the whole sequence instead of sequential input one by one.
Transformer uses self-attention blocks as well as feed-forward neural networks to extract the correlation between amino acids in the peptide sequence.
After the peptide representation is extracted by one of the above sequence models, it is resized into a 1d vector for the decoder. 
We uniformly apply a Multi-Layer Perceptron (MLP) \cite{lecun2015deep} decoder to predict the AP value of the peptide.
We use mean square error (MSE) loss to reduce the error between the predicted AP value and the ground-truth value.
As training progresses, the model is gradually transformed from underfitting to overfitting.
We select the model state that performs best on the validation set as the adequately fitted model.

\subsection{Encoding Peptides in Molecular Graphs}\label{4.2}

In the CGMD, we have coarsely grained the peptide molecules so that each amino acid consists of 1 to 5 beads.
A peptide of 5 to 10 amino acids can be represented as 5 to 50 beads with bonds, which is structural data.
Thus, GNNs, as deep learning methods for structural data, can be used to extract representations of peptides in CG molecular graphs.

The encoding and modelling of peptides in graphs are illustrated in Figure \ref{fig1}, indicated by the red lines.
We denote a peptide molecule as $\mathcal{G} = (\mathcal{V},\mathcal{E})$, where $\mathcal{V}=\{v_1,v_2,... ,v_n\}$ represents the beads and $\mathcal{E} \subseteq {\mathcal{V}}\times{\mathcal{V}}$ represents the existence of chemical bonds between the beads.
The edge set $\mathcal{E}$ can be represented as a binary adjacent matrix $A \in \{0,1\}^{N\times N}$, where $A_{i,j}=1$ if $(i,j)\subseteq \mathcal{E}$.
It is noteworthy that in the adjacency matrix of peptides, the connections between amino acids are represented by the links between backbone beads, and all other links are distributed within the amino acids.
Twenty types of amino acids can be represented by 15 types of beads.
We adopt the input embedding, the same as in the sequence model, to map $N$ discrete beads onto a feature matrix $X \subseteq N\times D$ ($D$: the dimension of the input embedding features) on the continuous space.
Thus, a peptide molecular graph is represented as the features of beads and the connections between beads, i.e., $\mathcal{G} = (A,X)$.

Predicting the AP value of a peptide molecular graph is a graph-level regression task.
We use GCN \cite{kipf2016semi}, GAT \cite{velivckovic2017graph}, and GraphSAGE \cite{hamilton2017inductive} to embed the graphs and extract graph-level representations through the readout operation.
As a spectral-based graph convolutional network, GCN has a solid mathematical foundation in graph signal processing \cite{sandryhaila2013discrete}.
The aggregation mechanism of GCN can be demonstrated as averaging the features of the ego and neighbouring nodes.
In contrast to GCN's adjacent matrix with edge weights of 0 or 1, GAT uses a learnable attention coefficient to control the share of information from each neighbouring node.
GAT learns the importance of edges between nodes compared to GCN, making the information aggregation mechanism soft and learnable.
GraphSAGE incorporates dropout \cite{srivastava2014dropout} into the information aggregation mechanism.
In the training phase, some of the neighbours of the node are ignored, and only a part of the neighbours participate in the feature convolution.

We then apply a readout to the node-level representations to obtain a graph-level representation of the peptide molecular graph.
The expression for the readout function is $h_{\mathcal{G}}=Readout(h_i|\forall v_i\in \mathcal{V})$,
where $h_{\mathcal{G}}$ denotes the graph-level representation of the peptide and $Readout(\cdot)$ denotes a readout function.
We uniformly use mean pooling as the readout function.
The graph-level representation is expressed as a one-dimensional vector.
For the decoder part, we use an MLP to predict the AP values of the peptides.
The MSE loss is used as the loss function to optimize the model parameters.
The model state that performs best on the validation set is considered to be a well-fitted model.

\subsection{Discussion on Learning Peptide Representation by Sequence-based and Graph-based Models} \label{4.3}

We adopt graph representation and sequence representation to investigate the effect of peptide structure representation approaches on model accuracy. 
In graph representation, peptides with various properties (i.e., polarity, charge, hydrogen donating and accepting capability, ring type with $\pi$-interaction) are directly input as features, whereas in sequence representation, only the letters denoting amino acid types are input, and the aforementioned properties are hidden and have to be captured by the deep learning algorithms. 
From the perspective of bioinformatic richness, graph encoding contains more information than sequence encoding.
However, from a mathematical point of view, graph and sequence encoding contain equivalent information because the sequence encoding is bijective to the graph encoding.
For applications with cyclic peptides, sequence encoding requires a defined permutation, which does not ensure model invariance to the starting amino acid.
In contrast, graph encoding can represent cyclic peptides as only the structure is expressed without the permutation.
Computationally, the sequence- and graph-based frameworks differ as well.
The inference efficiency of sequence- and graph-based models is similar.
Recurrent models, despite the need to iteratively input amino acids, the light weight of the models makes the inference faster than the parallel processing of Transformer and GNNs.
During data processing, transforming or reading graph-encoded peptides leads to additional time costs.
Overall, the advantage of graph-based frameworks is their ability to encode a larger peptide chemical space, including cyclic peptides, while sequence-based frameworks have a slight advantage in computational efficiency.

\section{Experiments}

\subsection{Dataset Statistics}
\begin{figure}
    \centering
    \includegraphics[width=\hsize]{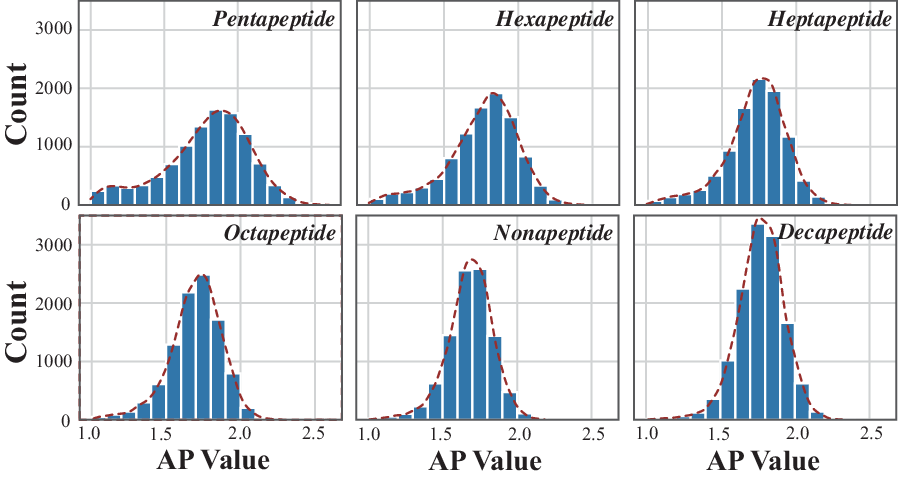}
    \caption{The label (AP value) distribution of the dataset.}
    \label{statistics}
\end{figure}

The self-assembly dataset contains peptides from pentapeptides to decapeptides. 
The dataset contains approximately 10,000 peptides of each length, giving a total of 62,159 peptide samples.
The distribution of AP values for each peptide length is shown in Figure \ref{statistics}.
Figure \ref{statistics} is a bar chart where the horizontal coordinate represents the AP value intervals, and the vertical coordinate represents the frequency of peptides distributed in the corresponding interval.
The distribution of the AP values of the peptides is close to a Gaussian distribution.
Comparing the distribution of AP values from pentapeptides to decapeptides, it is observable that the distribution of longer peptides tends to have a lower standard deviation. 
In the distribution of longer peptides, more samples are distributed in the interval around the mean value.
The vast majority of the peptides are distributed in the interval close to the mean and in the interval with low AP values.
The peptides with high AP values represent a small percentage of the chemical space.

\subsection{Benchmark Models}
We present benchmark models based on three encoding options on the self-assembled dataset.
Models based on amino acid sequence encoding include RNN \cite{elman1990rnn}, LSTM \cite{hochreiter1997lstm}, Bi-LSTM \cite{schuster1997bilstm}, and Transformer \cite{vaswani2017attention}.
Models based on molecular graph encoding include GCN \cite{kipf2016semi}, GAT \cite{velivckovic2017graph}, and GraphSAGE \cite{hamilton2017inductive}.
The above encoding schemes and benchmark models have been elaborated in Sections \ref{4.1} and \ref{4.2}, and the formulas for these deep learning models are provided in the supplementary.
These encoder models uniformly use the MLP as the decoder.
The MLP decoder contains four layers, each containing a fully connected layer, a non-linear activation function LeakyReLU, and batch normalization.
In addition, we introduce a baseline based on 1D vector encoding with Support Vector Machine (SVM) \cite{hearst1998support}, Random Forest (RF) \cite{breiman2001random} and MLP \cite{lecun2015deep}.
The baseline directly concatenates the word embedding features of the amino acids into a 1D vector as the representation of the peptide sequence and uses MLP to predict the AP value.

\subsection{Experimental Setup}

Out of 62,159 peptide samples, 4,000 peptides are randomly selected as the test set. 
The results shown in the experiments are uniformly the performance of the models on the test set. 
The remaining samples are divided into a training set and a validation set. 
The validation set contains another 4,000 peptides, and the training set contains 54,159 peptides. 
The benchmark models are trained with the training set, and the fitted models are selected based on the performance of the validation set.

The dataset contains a regression task and a classification task. The original labels are expressed as AP values on a continuous space, which is a regression task. Based on the median value, the dataset is divided into two parts with low and high AP values, which are labelled into two classes. 
We exclude samples with AP values very close to the median value, as they are predicted to be inappropriate for both classes.
The metrics used to evaluate the performance of the benchmark models in the regression task are Mean Absolute Error (MAE), Mean Square Error (MSE), and R-square (R$^2$); the metrics used to evaluate the performance of the benchmark models in the classification task are Accuracy (Acc), Precision, Recall, and F1-score.

\subsection{Evaluations}
In this section, we present and analyse the performance of the benchmark models on the peptide self-assembly dataset.
From the experimental results, we focus on the competition between sequence encoding and graph encoding.
We compare the performance of the encoding approaches by comparing the popular modelling approaches under each encoding method.
Taking into account the applicability and complexity of each encoding approach analysed in Section \ref{4.3}, we compare the suitability of each encoding approach on self-assembly as well as peptide-related tasks in an integrated manner.

\subsubsection{Performance on Regression Task}

\begin{table}[]
\centering
\caption{Performance of the benchmark models on the AP regression task. The best-performed results on each metric are bolded, and the second-ranked results are underlined.}
\label{table1}
\setlength{\tabcolsep}{2.35mm}
\begin{tabular}{llccc}
\hline
\multicolumn{1}{c}{\textBF{Encoding}} & \multicolumn{1}{l}{\textBF{Model}} & \multicolumn{1}{l}{\textBF{MAE}} & \multicolumn{1}{l}{\textBF{MSE}} & \multicolumn{1}{l}{\textBF{R$^2$}} \\ \hline
\multirow{4}{*}{1D vector} 
& RF                                & 0.0951                           & 0.01490                          & 0.662                           \\ 
& Batra et al. \cite{batra2022machine}                               & 0.0633                           & 0.00632                          & 0.847                           \\ 
& SVM                                & 0.0626                           & 0.00639                          & 0.857                         \\ 
& MLP                                & 0.0465                           & 0.00357                          & 0.920                           \\ \hline
\multirow{4}{*}{Sequence}                    & RNN                                & 0.0449                           & 0.00318                          & 0.929                           \\
& LSTM                               & 0.0422                           & 0.00281                          & 0.937                           \\
& Bi-LSTM                            & 0.0433                           & 0.00299                          & 0.933                           \\
& Transformer                        & \textBF{0.0391}                           & \textBF{0.00248}                          & \textBF{0.944}                           \\ \hline
\multirow{3}{*}{Graph}                     & GCN                                & 0.0440                           & 0.00315                          & 0.930                           \\
& GAT                                & 0.0422                           & 0.00290                          & 0.935                           \\
& GraphSAGE                          & \underline{0.0397}                           & \underline{0.00252}                          & \textBF{0.944}                           \\ \hline
\end{tabular}
\end{table}

The performance of the encoding methods with the benchmark models on the AP value regression task is shown in Table \ref{table1}.
The performance of Batra et al. based on RF is referenced from the results shown in the original paper \cite{batra2022machine}, as its implementation is not open source.
In the 1D vector embedding group, the deep learning-based MLP outperforms the ML approaches.
Both graph and sequence encoding significantly outperform the 1D vector encoding as the baseline among the three encoding methods.
Of the modelling approaches for sequence encoding, Transformer performs the best among the four benchmark models.
Transformer achieves 0.0391, 0.00248, and 0.944 for MAE, MSE, and R$^2$, respectively.
Among the sequence modelling methods other than Transformer, LSTM performs better than Bi-LSTM, while RNN performs worst.
Among the modelling methods for graph encoding, the models are ranked from best to worst performance as GraphSAGE, GAT, and GCN.
GraphSAGE achieves 0.0397, 0.00252, and 0.944 for MAE, MSE, and R$^2$, respectively.
Comparing the modelling methods of sequence encoding and graph encoding, Transformer is slightly better than GraphSAGE, and the performance of both is very close.
The performance of GAT and LSTM is similar, and GCN performs better than RNN and is second to Bi-LSTM.

The best performing model under sequential encoding is Transformer. The advantage of Transformer over vanilla RNN and the RNN-based LSTM and Bi-LSTM can be attributed to the self-attention block. The self-attention block evenly extracts the correlation between amino acids in the peptide, whereas RNNs process amino acids recurrently, resulting in long-term relationships that are challenging to capture. In addition, the self-attention block allows the Transformer to include more parameters, which is beneficial for learning more complex data-label correlations. The best performing model with graphical encoding is GraphSAGE, which, along with GCN and GAT, is a GNN algorithm based on feature aggregation. GraphSAGE employs a neighbour sampling approach, which randomly samples a subgraph for a node during each training epoch and extracts the neighbours' information in the subgraph through an aggregation function. As a result, the training of GraphSAGE can benefit from data augmentation. However, the GCN and GAT algorithms are both trained on the original graph without augmentation, which explains the superiority of GraphSAGE in terms of performance.

\begin{figure}
    \centering
    \includegraphics[width=\hsize]{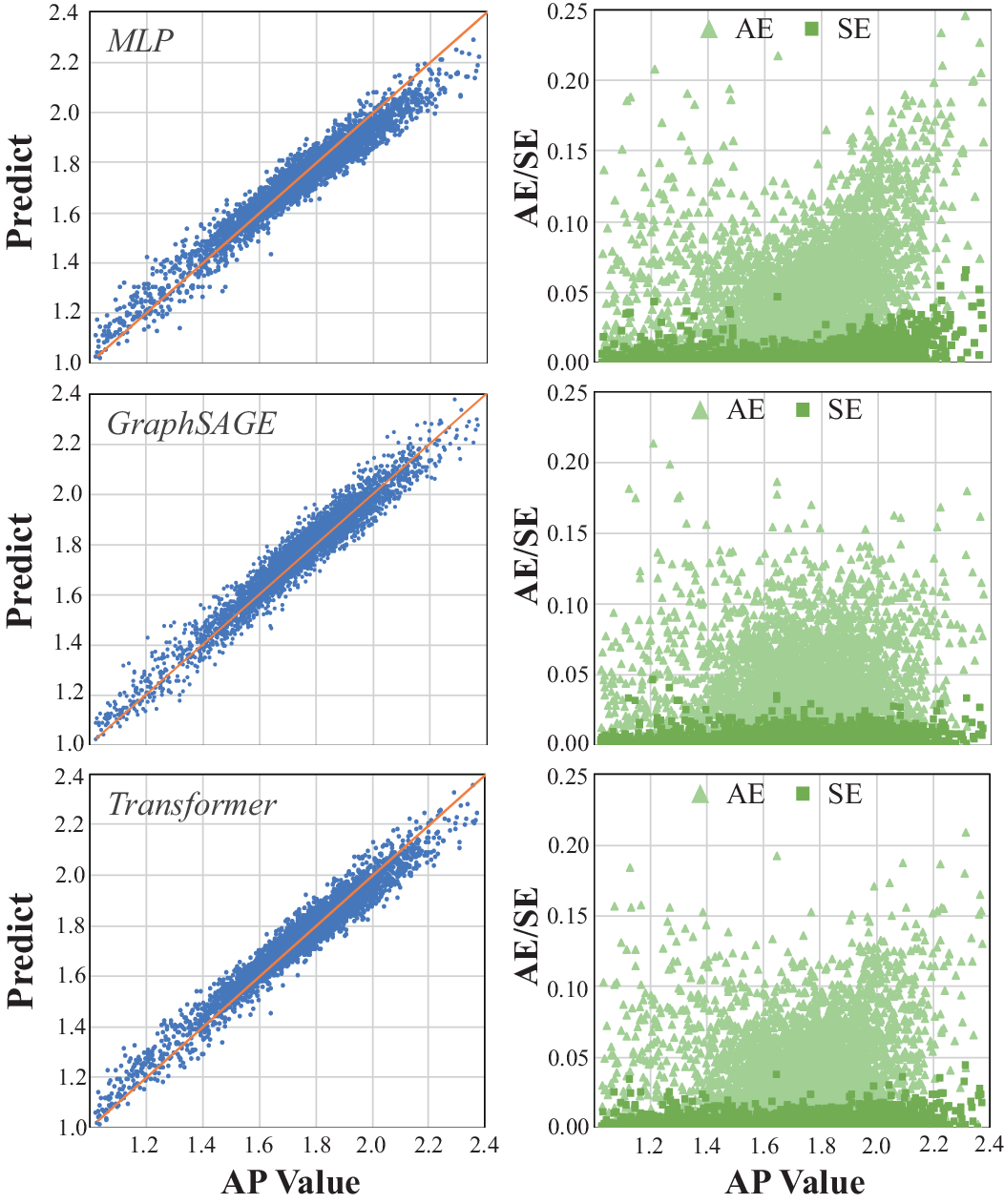}
    \caption{Visualisation of AP predictions. Left: Scatter plot of the predictions against their true AP values. The red auxiliary line indicates an error of 0. Right: distribution of absolute/squared error with respect to AP value.}
    \label{fig3}
\end{figure}

In Figure \ref{fig3}, we visualise the distribution of the model's predicted AP values around the ground-truth values.
The visualised models include MLP, GraphSAGE, and Transformer.
For each model, we show the predicted AP values against ground-truth values (left) and the distribution of absolute error/squared error about the ground-truth AP value (right).
The red auxiliary line represents the ideal prediction with an error of 0.
We start by analysing the prediction of MLP.
In the low-AP interval (about 1.0-1.4), more peptides have higher predicted AP than their ground-truth AP; in the high-AP interval (about 1.8-2.4), more peptides have lower predicted AP than their ground-truth AP.
This phenomenon can be attributed to the distribution of peptide AP values shown in Figure \ref{statistics}.
As the high/low AP peptides are sparse, the model tends to give a peptide closer to the middle AP value interval as the predicted value.
From the distribution of errors (right), we can see that the model has relatively larger errors in the high/low AP value intervals.
We then compare the visualisation of GraphSAGE and Transformer.
GraphSAGE's AP predictions are relatively evenly distributed at both ends of the red auxiliary line.
This means that GraphSAGE's errors on the high/low AP value interval do not produce a significant trend.
Examining the Transformer's predictions, we find that Transformer tends to produce predictions above ground truth on the low AP value interval and slightly tends to produce predictions below the ground truth on the high AP interval.
In terms of the distribution of errors, both GraphSAGE and Transformer have lower errors relative to MLP on the high/low AP value intervals.
From the distribution of AP predictions, MLP and Transformer tend to produce higher predictions than ground truth on the low-AP interval. 
Both GraphSAGE and Transformer perform satisfactorily on the high-AP interval.
Transformer is slightly better than GraphSAGE in terms of metrics, but in practice, the performance of GraphSAGE and Transformer can be considered equal.

\subsubsection{Performance on Classification Task}

To evaluate the encodings in a classification task, we classify the self-assembly dataset into high-AP and low-AP based on the median value.
For the binary classification task, we use Precision, Recall and F1-score as metrics that capture the information in the confusion matrix.
The Precision represents the proportion of true positives (high-AP peptides) among the predicted positives.
The recall represents the proportion of true positives correctly predicted by the model among all high-AP peptides.
The F1-score is the harmonic mean of Precision and Recall.

Table \ref{table2} shows the performance of the benchmark models on the classification task.
First, all models have a Precision close to or equal to 1.
This means that almost all failed predictions are predictions of high-AP samples into the low-AP class because the peptides in the high-AP class are significantly fewer than those in the low-AP class.
From the results, Transformer and GraphSAGE achieve an accuracy of 0.949 and 0.948, respectively.
Transformer has a slightly higher F1-score since Transformer achieves a Precision of 1, which is slightly higher than GraphSAGE.
Among the other benchmark models, GCN and GAT for graph encoding perform better than MLP, while RNN, LSTM, and Bi-LSTM for sequence encoding perform worse than MLP.
According to the metrics, the best-performing of the benchmark models is Transformer, followed by GraphSAGE.

\begin{table}[]
\centering
\setlength{\tabcolsep}{2.62mm}
\caption{Performance of the benchmark models on the AP classification task.}
\label{table2}
\begin{tabular}{lcccc}
\hline
\textBF{Model} & \textBF{Acc} & \textBF{Precision} & \textBF{Recall} & \textBF{F1-score} \\ \hline
RF            & 0.791        & 0.983              & 0.655  & 0.786    \\
SVM           & 0.887        & 0.998              & 0.808  & 0.893    \\
MLP            & 0.932        & 0.999              & 0.885  & 0.939    \\ \hline
RNN            & 0.927        & 1.000              & 0.876  & 0.934    \\
LSTM           & 0.930        & 1.000              & 0.880  & 0.936    \\
BiLSTM         & 0.928        & 1.000              & 0.878  & 0.935    \\
Transformer    & \textBF{0.949}        & 1.000              & \textBF{0.913}  & \textBF{0.955}    \\ \hline
GCN            & 0.945        & 0.998              & 0.908  & 0.951    \\
GAT            & 0.941        & 0.999              & 0.900  & 0.947    \\
GraphSAGE      & \underline{0.948}        & 0.999              & \textBF{0.913}  & \underline{0.954}    \\ \hline
\end{tabular}
\end{table}

\section{Disscussion and Conclusion}

In this work, based on a large and accurate dataset of peptide self-assembly from CGMD simulations, we investigate the peptide encoding approach to the prediction accuracy of the self-assembly of oligopeptides. 
The various modelling approaches are based on sequence and graph encoding, respectively, and the former encodes a peptide as an amino acid sequence containing sequence information, while the latter encodes a peptide as a coarse-grained molecular graph. 
Peptide representations under both encoding techniques are extracted by a sequence/graph deep learning model and predicted using an MLP decoder.
Based on our evaluation of the benchmark deep learning models, the sequence encoding using Transformer to extract the peptide representations ranks first among all the benchmark models.
The graph encoding using GraphSAGE is the best-performed graph-based model, which has a very close performance to Transformer.
Both encoding methods outperform the models with a 1D vector encoding.
The high prediction accuracy achieved by both sequence encoding and graph encoding has the potential to be exploited in practical applications. 
However, sequence coding methods are limited to encoding the backbone of a sequential peptide.
Graph encoding has a relatively larger upper screening space than sequence encoding and is, therefore, potentially easier to find peptides that match the desired properties.
Therefore, there is a trade-off between screening space and computational accuracy when choosing between the sequence-based Transformer and the graph-based GraphSAGE for peptide encoding.
The best encoding technique found in this research provides a reference for encoding and modelling short peptides for predicting other properties, such as hydration free energy, LUMO-HOMO gap, isoelectric points, etc.
In addition, we provide the code and framework for predicting the AP of any peptides within the complete sequence space of oligopeptides with over ten trillion sequences. The peptide self-related applications, such as peptide hydrogels, semiconductors, and drug delivery mediums, can be expected to boost with more discoveries of self-assembling oligopeptides.
In addition, peptide self-assembly 
mechanisms involving various interactions such as hydrogen bonding and hydrophobicity will be analysed with the assistance of attribution analysis, potentially providing operational guidance on the selection of peptides with 
desired amino acids in experiments.
It should be noted that experimental verification is crucial to the success of all future methods in machine learning. This must be resolved by high throughput experimentation to increase the datasets available for implementing into deep learning models. The high-throughput experimental data on peptide self-assembly can be obtained by employing the autonomous peptide synthesis platform or by using the natural language processing algorithm to mine the available literature on peptide self-assembly, which we will leave to the broader readers in the field.

\section{Key Points}
\begin{itemize}
    \item This work firstly presents a peptide self-assembly simulation dataset. The dataset provides more than 60,000 oligopeptides with their aggregation propensity (AP) as the evaluation indicator. We use the Martini force field \cite{marrink2007martini,monticelli2008martini} for coarse-grained molecular dynamics simulation. The validity of the simulation method on self-assembly task has been verified by previous self-assembly work \cite{frederix2015exploring,batra2022machine}.
    Compared to previous work, this paper focuses on peptides with lengths of 5 to 10.
    \item Compared with the experimental data, the simulation dataset is larger and less noisy, so we use deep learning to predict the AP of peptides. We notice that peptides could be encoded by either a sequential deep learning model or a graphical one according to different representations (amino acid sequence/molecular graph). So far we haven't found a benchmark work that discusses the advantages of different representations.
    \item We use the most general and advanced sequence and graph models to conduct a fair and broad benchmarking evaluation to find out how each representation is embedded in deep learning in the current deep learning environment. Extensive experiments have demonstrated that peptides based on amino acid sequence representation have the best performance with Transformer encoder, followed by GraphSAGE encoder which encodes peptides into molecular graphs.
\end{itemize}

\section{Data Avaliable}
The dataset and deep learning models are open-sourced at \url{https://github.com/Zihan-Liu-00/DL_for_Peptide}.

\section{Supplementary File}

Supplementary File has been submitted.

\section{Biographical Note}

\textbf{Mr. Zihan Liu} obtained master from the University of Edinburgh and now a Ph.D. candidate in Zhejiang University \& Westlake University. His research interest is explainable AI and AI for science.

\noindent\textbf{Dr. Jiaqi Wang} obtained Ph.D. from the University of Tennessee and now is working as a research assistant professor in Westlake University. His research interest is coupling machine learning for various biological applications. 

\noindent\textbf{Mr. Yun Luo} obtained bachelor from Wuhan University and now a Ph.D. candidate in Zhejiang University \& Westlake University. His research interest is natural language processing and graph neural network.

\noindent\textbf{Miss. Shuang Zhao} obtained master from Imperial College London and now a Ph.D. student in Tsinghua University, conducting research on biological and optical system.   

\noindent\textbf{Dr. Wenbin Li} obtained Ph.D. from MIT and conducted postdoctoral research in Oxford. Now he is an assistant professor in Westlake University, focusing on molecular simulations. 

\noindent\textbf{Prof. Stan Z. Li}, IEEE Fellow. He obtained Ph.D. from Surrey University.
He was the director of the Center for Biometrics and Security Research (CBSR), Chinese Academy of Sciences. 
Now He is a Chair Professor of Artificial Intelligence in Westlake University.

\section{Acknowledgments}
Z.L. and S.Z.L. are supported by National Key R\&D Program of China (No. 2022ZD0115100), National Natural Science Foundation of China (No. U21A20427), and Project (No. WU2022A0XX) from the Center of Synthetic Biology and Integrated Bioengineering of Westlake University.
J.W., S.Z., and W.L. are supported by Research Center for Industries of the Future at Westlake University under Award No. WU2022C041. J.W. is also supported by Zhejiang Postdoctoral Science Foundation (No. 103346582102) and National Natural Science Foundation of China (No. 52101023).

\bibliographystyle{unsrt}
\bibliography{ref}

\begin{thebibliography}{10}

\bibitem{langel2009introduction}
Ulo Langel, Benjamin~F Cravatt, Astrid Graslund, NGH Von~Heijne, Matjaz Zorko,
  Tiit Land, and Sherry Niessen.
\newblock {\em Introduction to peptides and proteins}.
\newblock CRC press, 2009.

\bibitem{cinar2012amyloid}
Goksu Cinar, Hakan Ceylan, Mustafa Urel, Turan~S Erkal, E~Deniz~Tekin, Ayse~B
  Tekinay, Aykutlu D$\hat{\mathbf{a}}$na, and Mustafa~O Guler.
\newblock Amyloid inspired self-assembled peptide nanofibers.
\newblock {\em Biomacromolecules}, 13(10):3377--3387, 2012.

\bibitem{fan2018near}
Zhen Fan, Yan Chang, Chaochu Cui, Leming Sun, David~H Wang, Zui Pan, and
  Mingjun Zhang.
\newblock Near infrared fluorescent peptide nanoparticles for enhancing
  esophageal cancer therapeutic efficacy.
\newblock {\em Nature Communications}, 9(1):1--11, 2018.

\bibitem{tao2017self}
Kai Tao, Pandeeswar Makam, Ruth Aizen, and Ehud Gazit.
\newblock Self-assembling peptide semiconductors.
\newblock {\em Science}, 358(6365):eaam9756, 2017.

\bibitem{zhao2010molecular}
Xiubo Zhao, Fang Pan, Hai Xu, Mohammed Yaseen, Honghong Shan, Charlotte~AE
  Hauser, Shuguang Zhang, and Jian~R Lu.
\newblock Molecular self-assembly and applications of designer peptide
  amphiphiles.
\newblock {\em Chemical Society Reviews}, 39(9):3480--3498, 2010.

\bibitem{levin2020biomimetic}
Aviad Levin, Tuuli~A Hakala, Lee Schnaider, Gon{\c{c}}alo~JL Bernardes, Ehud
  Gazit, and Tuomas~PJ Knowles.
\newblock Biomimetic peptide self-assembly for functional materials.
\newblock {\em Nature Reviews Chemistry}, 4(11):615--634, 2020.

\bibitem{whitesides2002self}
George~M Whitesides and Bartosz Grzybowski.
\newblock Self-assembly at all scales.
\newblock {\em Science}, 295(5564):2418--2421, 2002.

\bibitem{krause2014steering}
Matthias Krause and Alexis Gautreau.
\newblock Steering cell migration: lamellipodium dynamics and the regulation of
  directional persistence.
\newblock {\em Nature Reviews Molecular Cell Biology}, 15(9):577--590, 2014.

\bibitem{aguzzi2009prions}
Adriano Aguzzi and Anna~Maria Calella.
\newblock Prions: protein aggregation and infectious diseases.
\newblock {\em Physiological Reviews}, 89(4):1105--1152, 2009.

\bibitem{knowles2014amyloid}
Tuomas~PJ Knowles, Michele Vendruscolo, and Christopher~M Dobson.
\newblock The amyloid state and its association with protein misfolding
  diseases.
\newblock {\em Nature Reviews Molecular Cell Biology}, 15(6):384--396, 2014.

\bibitem{chiti2017protein}
Fabrizio Chiti and Christopher~M Dobson.
\newblock Protein misfolding, amyloid formation, and human disease: a summary
  of progress over the last decade.
\newblock {\em Annu. Rev. Biochem}, 86(1):27--68, 2017.

\bibitem{rudd1998coarse}
Robert~E Rudd and Jeremy~Q Broughton.
\newblock Coarse-grained molecular dynamics and the atomic limit of finite
  elements.
\newblock {\em Physical Review B}, 58(10):R5893, 1998.

\bibitem{frederix2015exploring}
Pim~WJM Frederix, Gary~G Scott, Yousef~M Abul-Haija, Daniela Kalafatovic,
  Charalampos~G Pappas, Nadeem Javid, Neil~T Hunt, Rein~V Ulijn, and Tell
  Tuttle.
\newblock Exploring the sequence space for (tri-) peptide self-assembly to
  design and discover new hydrogels.
\newblock {\em Nature Chemistry}, 7(1):30--37, 2015.

\bibitem{batra2022machine}
Rohit Batra, Troy~D Loeffler, Henry Chan, Srilok Srinivasan, Honggang Cui,
  Ivan~V Korendovych, Vikas Nanda, Liam~C Palmer, Lee~A Solomon, H~Christopher
  Fry, et~al.
\newblock Machine learning overcomes human bias in the discovery of
  self-assembling peptides.
\newblock {\em Nature Chemistry}, 14(12):1427--1435, 2022.

\bibitem{zapadka2017factors}
Karolina~L Zapadka, Frederik~J Becher, AL~Gomes~dos Santos, and Sophie~E
  Jackson.
\newblock Factors affecting the physical stability (aggregation) of peptide
  therapeutics.
\newblock {\em Interface Focus}, 7(6):20170030, 2017.

\bibitem{marullo2013peptide}
Rachel Marullo, Mark Kastantin, Laurie~B Drews, and Matthew Tirrell.
\newblock Peptide contour length determines equilibrium secondary structure in
  protein-analogous micelles.
\newblock {\em Biopolymers: Original Research on Biomolecules}, 99(9):573--581,
  2013.

\bibitem{seebach2006helices}
Dieter Seebach, David~F Hook, and Alice Gl{\"a}ttli.
\newblock Helices and other secondary structures of $\beta$-and
  $\gamma$-peptides.
\newblock {\em Peptide Science: Original Research on Biomolecules},
  84(1):23--37, 2006.

\bibitem{hochreiter1997lstm}
Sepp Hochreiter and J{\"u}rgen Schmidhuber.
\newblock Long short-term memory.
\newblock {\em Neural Computation}, 9(8):1735--1780, 1997.

\bibitem{vaswani2017attention}
Ashish Vaswani, Noam Shazeer, Niki Parmar, Jakob Uszkoreit, Llion Jones,
  Aidan~N Gomez, {\L}ukasz Kaiser, and Illia Polosukhin.
\newblock Attention is all you need.
\newblock {\em Advances in Neural Information Processing Systems}, 30, 2017.

\bibitem{kipf2016semi}
Thomas~N Kipf and Max Welling.
\newblock Semi-supervised classification with graph convolutional networks.
\newblock {\em arXiv preprint arXiv:1609.02907}, 2016.

\bibitem{velivckovic2017graph}
Petar Velickovic, Guillem Cucurull, Arantxa Casanova, Adriana Romero, Pietro
  Lio, Yoshua Bengio, et~al.
\newblock Graph attention networks.
\newblock {\em Stat}, 1050(20):10--48550, 2017.

\bibitem{hamilton2017inductive}
Will Hamilton, Zhitao Ying, and Jure Leskovec.
\newblock Inductive representation learning on large graphs.
\newblock {\em Advances in Neural Information Processing Systems}, 30, 2017.

\bibitem{mckay2000comparison}
Michael~D McKay, Richard~J Beckman, and William~J Conover.
\newblock A comparison of three methods for selecting values of input variables
  in the analysis of output from a computer code.
\newblock {\em Technometrics}, 42(1):55--61, 2000.

\bibitem{abraham2015gromacs}
Mark~James Abraham, Teemu Murtola, Roland Schulz, Szil{\'a}rd P{\'a}ll,
  Jeremy~C Smith, Berk Hess, and Erik Lindahl.
\newblock Gromacs: High performance molecular simulations through multi-level
  parallelism from laptops to supercomputers.
\newblock {\em SoftwareX}, 1:19--25, 2015.

\bibitem{marrink2007martini}
Siewert~J Marrink, H~Jelger Risselada, Serge Yefimov, D~Peter Tieleman, and
  Alex~H De~Vries.
\newblock The martini force field: coarse grained model for biomolecular
  simulations.
\newblock {\em The Journal of Physical Chemistry B}, 111(27):7812--7824, 2007.

\bibitem{monticelli2008martini}
Luca Monticelli, Senthil~K Kandasamy, Xavier Periole, Ronald~G Larson, D~Peter
  Tieleman, and Siewert-Jan Marrink.
\newblock The martini coarse-grained force field: extension to proteins.
\newblock {\em Journal of Chemical Theory and Computation}, 4(5):819--834,
  2008.

\bibitem{huang2013charmm36}
Jing Huang and Alexander~D MacKerell~Jr.
\newblock Charmm36 all-atom additive protein force field: Validation based on
  comparison to nmr data.
\newblock {\em Journal of Computational Chemistry}, 34(25):2135--2145, 2013.

\bibitem{lee2012modeling}
One-Sun Lee, Vince Cho, and George~C Schatz.
\newblock Modeling the self-assembly of peptide amphiphiles into fibers using
  coarse-grained molecular dynamics.
\newblock {\em Nano Letters}, 12(9):4907--4913, 2012.

\bibitem{xiong2019conformation}
Qinsi Xiong, Yixiang Jiang, Xiang Cai, Fadeng Yang, Zigang Li, and Wei Han.
\newblock Conformation dependence of diphenylalanine self-assembly structures
  and dynamics: Insights from hybrid-resolution simulations.
\newblock {\em ACS Nano}, 13(4):4455--4468, 2019.

\bibitem{mueller2016machine}
Tim Mueller, Aaron~Gilad Kusne, and Rampi Ramprasad.
\newblock Machine learning in materials science: Recent progress and emerging
  applications.
\newblock {\em Reviews in Computational Chemistry}, 29:186--273, 2016.

\bibitem{halevy2009unreasonable}
Alon Halevy, Peter Norvig, and Fernando Pereira.
\newblock The unreasonable effectiveness of data.
\newblock {\em IEEE intelligent systems}, 24(2):8--12, 2009.

\bibitem{nikolenko2021synthetic}
Sergey~I Nikolenko.
\newblock {\em Synthetic data for deep learning}, volume 174.
\newblock Springer, 2021.

\bibitem{tremblay2018training}
Jonathan Tremblay, Aayush Prakash, David Acuna, Mark Brophy, Varun Jampani, Cem
  Anil, Thang To, Eric Cameracci, Shaad Boochoon, and Stan Birchfield.
\newblock Training deep networks with synthetic data: Bridging the reality gap
  by domain randomization.
\newblock In {\em Proceedings of the IEEE conference on computer vision and
  pattern recognition workshops}, pages 969--977, 2018.

\bibitem{elman1990rnn}
Jeffrey~L Elman.
\newblock Finding structure in time.
\newblock {\em Cognitive Science}, 14(2):179--211, 1990.

\bibitem{schuster1997bilstm}
Mike Schuster and Kuldip~K Paliwal.
\newblock Bidirectional recurrent neural networks.
\newblock {\em IEEE Transactions on Signal Processing}, 45(11):2673--2681,
  1997.

\bibitem{lecun2015deep}
Yann LeCun, Yoshua Bengio, and Geoffrey Hinton.
\newblock Deep learning.
\newblock {\em Nature}, 521(7553):436--444, 2015.

\bibitem{sandryhaila2013discrete}
Aliaksei Sandryhaila and Jos{\'e}~MF Moura.
\newblock Discrete signal processing on graphs.
\newblock {\em IEEE Transactions on Signal Processing}, 61(7):1644--1656, 2013.

\bibitem{srivastava2014dropout}
Nitish Srivastava, Geoffrey Hinton, Alex Krizhevsky, Ilya Sutskever, and Ruslan
  Salakhutdinov.
\newblock Dropout: a simple way to prevent neural networks from overfitting.
\newblock {\em The Journal of Machine Learning Research}, 15(1):1929--1958,
  2014.

\bibitem{hearst1998support}
Marti~A. Hearst, Susan~T Dumais, Edgar Osuna, John Platt, and Bernhard
  Scholkopf.
\newblock Support vector machines.
\newblock {\em IEEE Intelligent Systems and Their Applications}, 13(4):18--28,
  1998.

\bibitem{breiman2001random}
Leo Breiman.
\newblock Random forests.
\newblock {\em Machine Learning}, 45:5--32, 2001.

\end{thebibliography}

\end{document}